\documentclass[epjCONF]{svjour}
\usepackage{graphicx}
\usepackage[varg]{txfonts} 
\usepackage[latin1]{inputenc}
\session-title{The Space Photometry Revolution}
\begin{document}
\title{Understanding tidal dissipation in gaseous giant planets from their core to their surface}
\author{M. Guenel\inst{1}\fnmsep\thanks{\email{mathieu.guenel@cea.fr}} \and S. Mathis\inst{1}\fnmsep\thanks{\email{stephane.mathis@cea.fr}} \and F. Remus\inst{2,1}\fnmsep\thanks{\email{francoise.remus@obspm.fr}} }
\institute{Laboratoire AIM Paris-Saclay, CEA/DSM/IRFU/SAp - Universit\'e Paris Diderot - CNRS, 91191 Gif-sur-Yvette, France \and IMCCE, Observatoire de Paris, CNRS UMR 8028, UPMC, USTL, 77 Avenue Denfert-Rochereau, 75014 Paris, France}
\abstract{
Tidal dissipation in planetary interiors is one of the key physical mechanisms that drive the evolution of star-planet and planet-moon systems. Tidal dissipation in planets is intrinsically related to their internal structure. In particular, fluid and solid layers behave differently under tidal forcing. Therefore, their respective dissipation reservoirs have to be compared. In this work, we compute separately the contributions of the potential dense rocky/icy core and of the convective fluid envelope of gaseous giant planets, as a function of core size and mass. We demonstrate that in general both mechanisms must be taken into account.
} 
\maketitle

\section{Modelling the tidal dissipation mechanisms}
\label{sec:modelling}
\begin{figure}[!ht]
\centering
\includegraphics[width=0.5\textwidth]{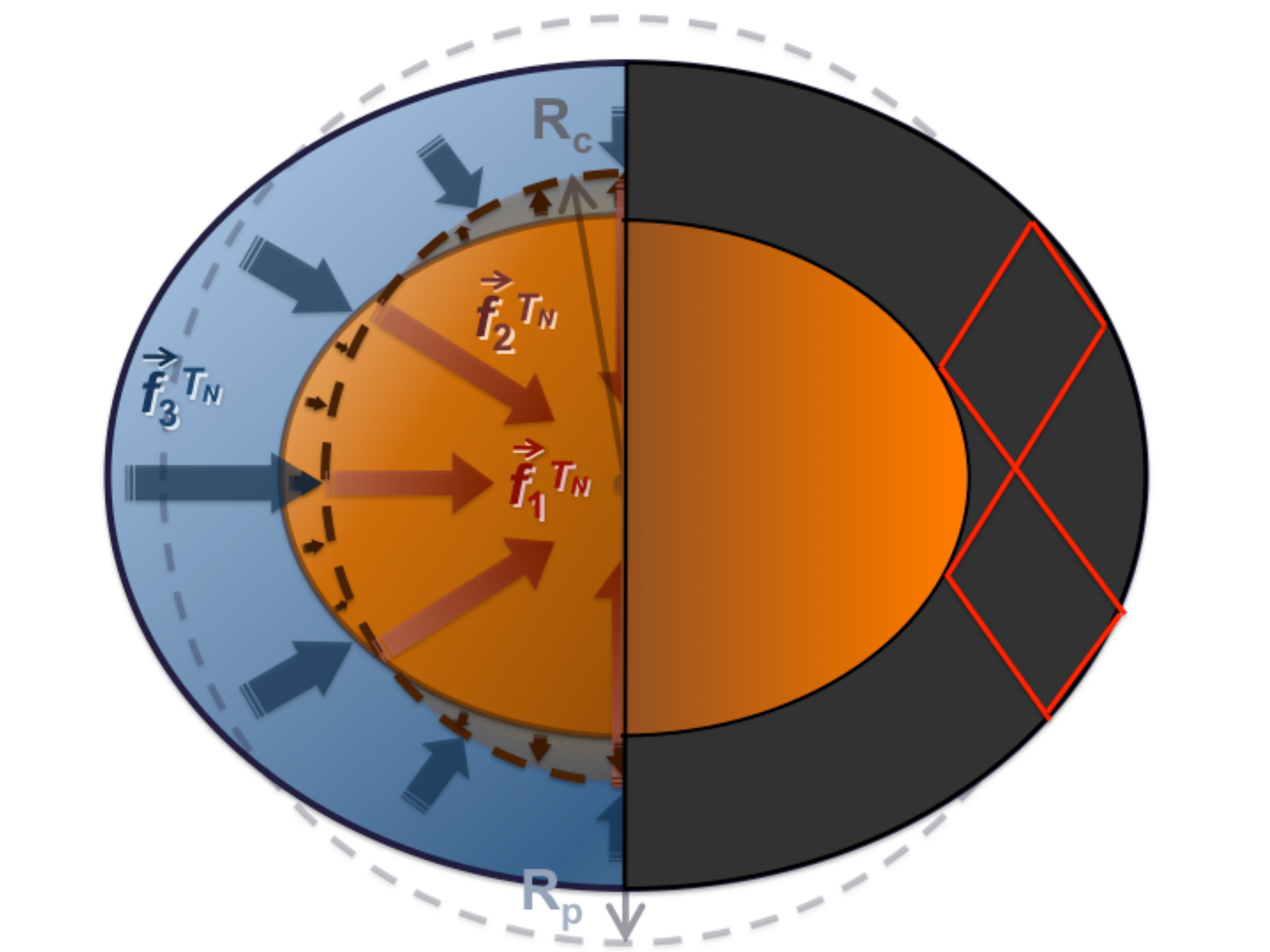}
\caption{{\bf Left~:} Gravitational forces ($\vec{f}_1$), internal constraints ($\vec{f}_2$) and hydrostatic pressure ($\vec{f}_3$) acting on the solid core, which is deformed by the tidal force exerted by the companion. {\bf Right~:} Attractor formed by a path of characteristics of inertial waves.}
\label{Guenel_M_fig1}
\end{figure}

We use a two-layer model that features a central planet A of mass $M_p$ and mean radius $R_p$ assumed to be in moderate solid-body rotation $\Omega$ with $\epsilon^2 \equiv \Omega^2/ \sqrt{\mathcal{G} M_p / R_p^3} \ll 1$\footnote{In this regime, the Coriolis acceleration, which scales as $\Omega$, is taken into account while the centrifugal acceleration, which scales as $\Omega^{2}$ is neglected.}. Its rocky (or icy) solid core of radius $R_c$ and density $\rho_c$ is surrounded by a convective fluid envelope of density $\rho_o$ --- both are assumed to be homogeneous for the sake of simplicity. A point-mass tidal perturber B of mass $M_B$ is orbiting A with a mean motion $n$.  The time-dependent tidal potential exerted by the companion leads to two different dissipation mechanisms illustrated by Fig. \ref{Guenel_M_fig1} :
\begin{itemize}
\item first, we consider the viscoelastic dissipation in the solid core, for which we assume that the rheology follows the linear rheological model of Maxwell with a rigidity $G$ and a viscosity $\eta$ ; we also assume that the surrounding envelope is inviscid and only applies hydrostatic pressure and gravitational attraction on the core;

\item then, the turbulent viscosity in the fluid convective envelope dissipates the kinetic energy of tidal inertial waves propagating in that region. The restoring force of inertial waves is the Coriolis acceleration and their frequency is smaller than the Coriolis frequency : $\omega \in [-2\Omega,2\Omega]$. Moreover, their kinetic energy may concentrate and form shear layers around attractor cycles, which leads to enhanced damping by turbulent viscosity. In order to compute the dissipation, the core is assumed to be perfectly rigid.
\end{itemize}

Using \cite{Remus2014,RMZL2012,Ogilvie2013}, we compute for each of these mechanisms the ``reservoir of dissipation", a weighted frequency-average of the imaginary part of the Love number $k^2_2(\omega) = {\Phi^2_2}'/U^2_2$ (which is the ratio at $r=R_p$ between the $Y^2_2$-components of the Eulerian perturbation $\Phi'$ of the self-potential of body A, and of the tidal potential $U$) defined as $\int_{-\infty}^{+\infty} \! {\rm Im} \left[k_2^2(\omega)\right] \,\frac{\mathrm{d}\omega}{\omega} = \int_{-\infty}^{+\infty} \! \left| k_2^2(\omega) \right|/Q_{2}^{2}(\omega) \,\frac{\mathrm{d}\omega}{\omega}$,
where $Q^2_2(\omega)$ is the corresponding tidal quality factor.

\section{Results}
\label{sec:results}
\begin{figure}[ht!]
\centering
\includegraphics[width=0.57\textwidth]{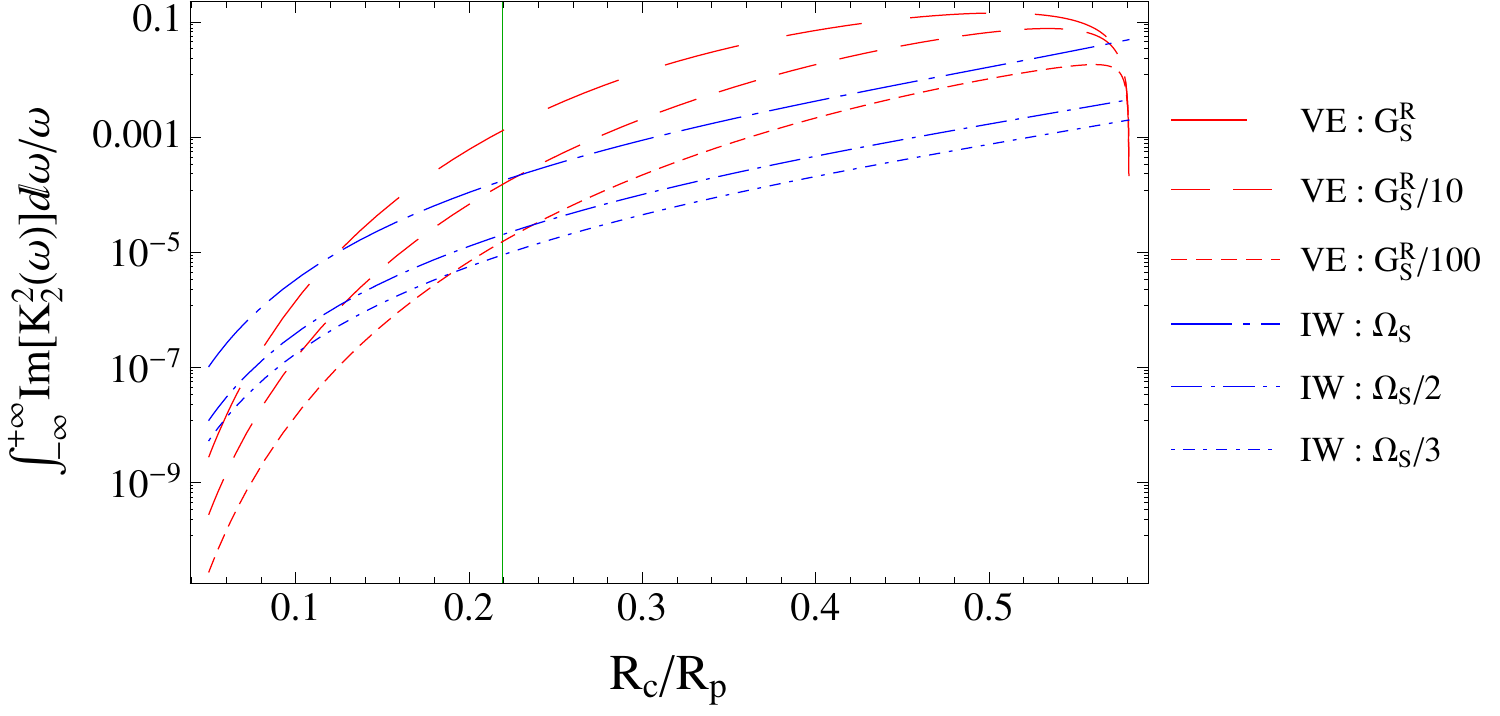}
\includegraphics[trim=25 0 100 0,clip,width=0.41\textwidth]{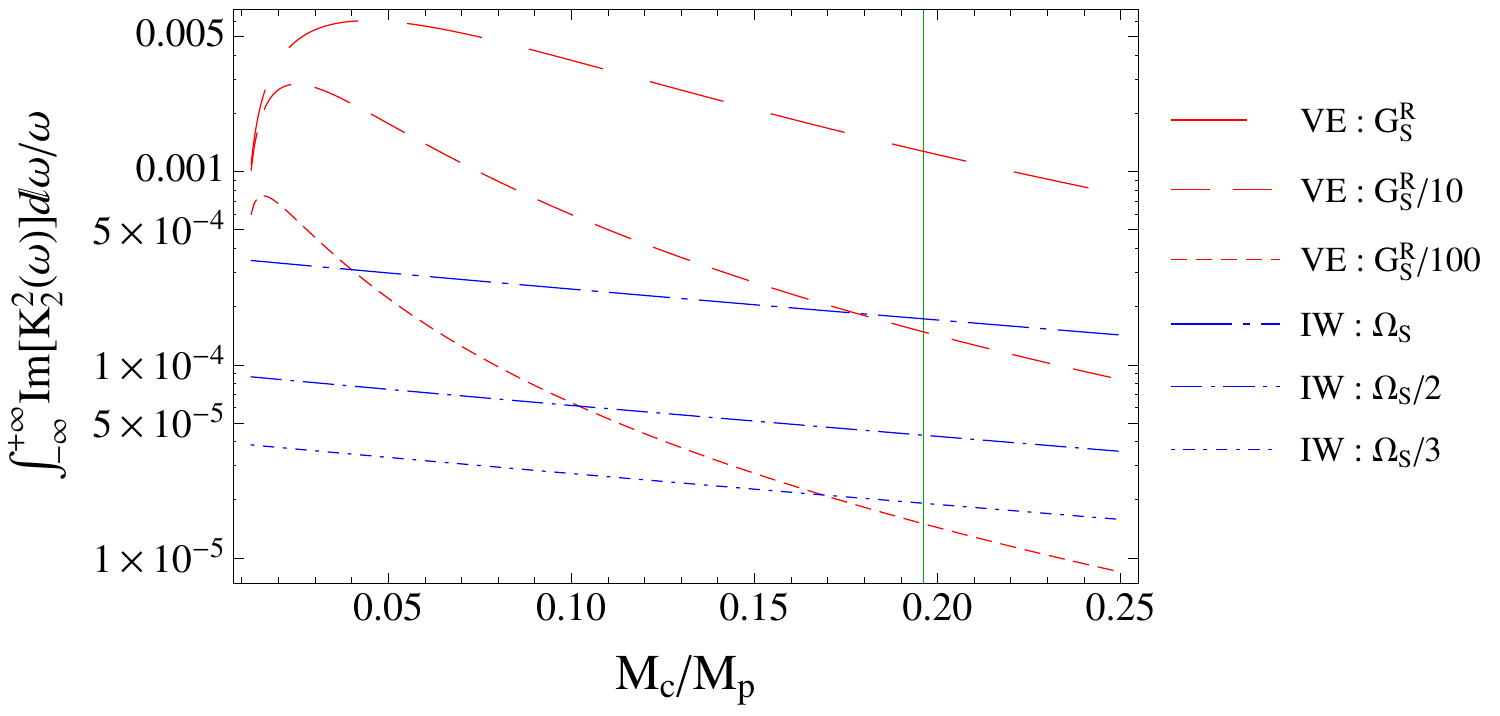}
\caption{{\bf Left~:} Dissipation reservoirs for the viscoelastic (VE) dissipation in the core (red curve) and the turbulent friction acting on inertial waves (IW) in the fluid envelope (blue curves) in Saturn-like planets as a function of $R_c/R_p$, $\Omega$, and $G$. We use $M_c/M_p = 0.196$. The vertical green line corresponds to $R_c/R_p = 0.219$. {\bf Right~:} Similar to the left-side but now as a function of $M_c/M_p$ with fixed $M_p$ and $R_p$. We adopt $R_c/R_p = 0.219$. The vertical green line corresponds to $M_c/M_p = 0.196$.}
\label{Guenel_M_fig2}
\end{figure}

These plots show that in Saturn-like gaseous giant planets, the two distinct mechanisms exposed earlier can both contribute to tidal dissipation, and that therefore none of them can be neglected in general (see \cite{GMR2014}).
This demonstrates that it is necessary to build complete models of tidal dissipation in planetary interiors from their centre to their surface without any arbitrary a-priori.

\bibliographystyle{epj}
\bibliography{Guenel_M}
%

\end{document}